\documentclass[aps,prc,letterpaper,11pt,amsmath,amssymb,twoside,nofootinbib,showpacs,preprint]{revtex4}
\usepackage{graphicx}
\usepackage{amsmath}
\usepackage{amsfonts}
\usepackage{amssymb}

\begin{document}
\newcommand{\be}{\begin{equation}}
\newcommand{\ee}{\end{equation}}
\newcommand{\bea}{\begin{eqnarray}}
\newcommand{\eea}{\end{eqnarray}}
\newcommand{\bef}{\begin{figure}}
\newcommand{\eef}{\end{figure}}
\newcommand{\bce}{\begin{center}}
\newcommand{\ece}{\end{center}}
\title{The $\bar{K}NN$ system with chiral dynamics}
\author{M. Bayar}
 \affiliation{Instituto de F{\'\i}sica Corpuscular (centro mixto CSIC-UV)\\
Institutos de Investigaci\'on de Paterna, Aptdo. 22085, 46071, Valencia, Spain}
\affiliation{Department of Physics, Kocaeli University, 41380 Izmit, Turkey}
\author{J. Yamagata-Sekihara}
\affiliation{Instituto de F{\'\i}sica Corpuscular (centro mixto CSIC-UV)\\
Institutos de Investigaci\'on de Paterna, Aptdo. 22085, 46071, Valencia, Spain}
\author{E. Oset}
\affiliation{Instituto de F{\'\i}sica Corpuscular (centro mixto CSIC-UV)\\
Institutos de Investigaci\'on de Paterna, Aptdo. 22085, 46071, Valencia, Spain}

\begin{abstract}

We have performed a calculation of the scattering amplitude for the three body 
system $\bar{K}NN$ assuming $\bar{K}$ scattering against a $NN$ cluster, using the Fixed 
Center approximation to the Faddeev equations. The $\bar{K}N$ amplitudes, which we take from
 chiral unitary dynamics, govern the reaction and we find a $\bar{K}NN$ amplitude that peaks around 40 MeV below 
the $\bar{K}NN$ threshold, with a width in $|T|^2$ of the order of $50$ MeV for spin 0 and has another peak around $27$ MeV with similar
width for spin 1. The results are in line with those obtained 
using different methods but implementing chiral dynamics. The simplicity of the approach allows one to see the important 
ingredients responsible for the results. In particular we show the effects from the reduction of the size of the $NN$
cluster due to the interaction with the $\bar{K}$ and those from the explicit considiration of the $ \pi \Sigma N $
channel in the three body equations.
\end{abstract}

\pacs{11.10.St; 12.40.Yx; 13.75.Jz; 14.20.Gk; 14.40.Df}

\vspace{1cm}

\date{\today}

\maketitle

\section{Introduction}
 \label{sec:intro}
 
 The study of the bound $\bar{K} NN$ system is rising much interest as the lightest nuclear system binding a $\bar{K}$ \cite{Ikeda:2007nz,Shevchenko:2006xy,Shevchenko:2007zz,Dote:2008in,Dote:2008hw,Ikeda:2008ub}. Earlier calculations on this system were already done in the 60's 
\cite{nogami} and more recently in \cite{akayama}, yet, the works of \cite{Ikeda:2007nz,Shevchenko:2006xy,Shevchenko:2007zz,Dote:2008in,Dote:2008hw,Ikeda:2008ub} improve considerably on those earlier works, and according to \cite{Shevchenko:2007zz}, the results of \cite{akayama} should be considered at best as a rough estimate.
   It is very instructive to recall the basics of these calculations. Those of  
\cite{Ikeda:2007nz,Shevchenko:2006xy,Shevchenko:2007zz,Ikeda:2008ub} use Faddeev equations, in the formulation of Alt-Grassberger-Sandhas 
(AGS) \cite{Alt:1967fx}, using a separable interaction for the potentials with energy independent strength, with form factors depending on 
the three momenta. They also include as coupled channels $ N \pi \Sigma$ and $NN \bar{K}$. On the other hand the works of 
 \cite{Dote:2008in,Dote:2008hw} use a variational method to obtain the binding energy and width, by means of an effective potential
 \cite{Hyodo:2007jq} that leads to the strongly energy dependent $\bar{K}N$ amplitudes of the chiral unitary approach 
\cite{Kaiser:1995eg,angels,ollerulf,Lutz:2001yb,Oset:2001cn,Hyodo:2002pk,cola,Borasoy:2005ie,Oller:2006jw,Borasoy:2006sr}.
 The works of \cite{Dote:2008in,Dote:2008hw} make the variational calculation using the $\bar{K}N$ effective potential. This includes
 the $\pi \Sigma$ channel integrated into the effective $\bar{K}N$ potential, but do not include the $N \pi \Sigma$ channel explicitly
 into the three body system. The $ \pi \Sigma$ two body channel
 is effectively considered through the $\bar{K}N$ potential, but the interaction of $\pi N$ or $\Sigma N$ which appears in the Faddeev 
equations would not be considered. Yet, the $\pi N$ interaction is relatively weak compared to the $\bar{K} N$ at these energies, and 
the effect of the $\Sigma N$ is very small as mentioned in \cite{Shevchenko:2007zz}, so one should look for other reasons to understand 
the differences between the two approaches. In Table IV of \cite{Shevchenko:2007zz} one finds that there are 11 MeV difference considering
just one channel $\bar{K}N$ or the two channels $\bar{K}N$ and $\pi \Sigma$, both in the evaluation of the $\bar{K}N$ amplitude and in the 
$\bar{K} NN-\pi \Sigma N$ system, with more binding in the case of two channels. One should note that this is not to be compared with the 
present approach or the one in \cite{Dote:2008in,Dote:2008hw} where the $\bar{K}N$ amplitude is always calculated with $\bar{K}N$, $\pi \Sigma$
and other coupled channels. As argued here and in \cite{Hyodo:2007jq} this should take into account most of the effects of the $\pi \Sigma$
channel. The extra Faddeev diagrams with explicit $\pi \Sigma N$ intermediate states involve the $\pi N$ or $\Sigma N$ interactions which 
have small effects. Actually, this line of argumentation coincides with a conjecture of  Schick and Gibson \cite{Schick:1978wi} which was
substantiated numerically in the work of Ref. \cite{Toker:1981zh}.

However, in  \cite{Dote:2008hw} the authors make guesses on the reasons for the differences between the Faddeev approach and their approach and,
among other reasons, they suggest that the three body $\pi \Sigma N$ dynamics not accounted for in their work could be in large part responsible
for the differences. In view of this persistent question we shall introduce here explicitly the $\pi \Sigma$ channel in the three body system
to find out an answer.

 The two Faddeev approaches lead to binding energies higher than the variational approach, 50-70 MeV versus around 20 MeV binding
 respectively. The detail mentioned above of an energy independent kernel used in the AGS equations is partly responsible for the extra
 binding of these approaches with respect to the chiral calculations. Indeed, as shown in \cite{wavefunction,wavejunko}, it is possible 
to obtain the same results as with the field theoretical chiral unitary approach, using Quantum Mechanics with a separable potential,
 but where the potential is energy dependent, proportional to the sum of the two external meson energies in $\bar{K}N \to \bar{K}N$.
 As a consequence, a smaller  $\bar{K}N \to \bar{K}N$ amplitude is obtained at lower $\bar{K} N$ energies, resulting in a smaller binding 
for the $\bar{K}NN$ system. This numerical result was already mentioned in \cite{Dote:2008hw}. Actually, the same result is found within 
the approach of \cite{Ikeda:2007nz,Ikeda:2008ub} when the energy dependence of the Weinberg-Tomozawa chiral potential is taken into account
 \cite{Ikeda:2010tk}. In return this latter approach contained two poles in the scattering matrices of the three body system, as a reminder
 of the two $\Lambda(1405)$ poles contained in \cite{cola,Borasoy:2005ie,Oller:2006jw,Borasoy:2006sr}. We shall come back to this work further
on.

Yet, in most cases the widths are 
systematically larger than the binding energy, of the order of 70-100 MeV. This certainly makes the observation of these states problematic, 
as acknowledged in all these works. In view of this, the claim in \cite{Agnello:2005qj} of a bound state of $K^- pp$ bound state with 115 MeV
 binding was met with skepticism, and soon it  was shown that the peak observed in 
\cite{Agnello:2005qj} was easily interpreted in terms of conventional, unavoidable, reaction mechanisms which were well under control 
\cite{Magas:2006fn}.

     In between, the study of three body systems involving mesons and baryons with the Faddeev equations has experienced a new qualitative jump. It is well known that different potentials leading to exactly the same on shell two body scattering amplitudes lead to different results in the Faddeev equations due to their different off shell extrapolation. Yet, a novel result has emerged from the study of the two meson-one baryon systems with chiral dynamics, since it was shown in  \cite{MartinezTorres:2007sr} that the off shell part of the two body scattering amplitudes cancels with genuine three body terms that the same chiral Lagrangians provide at the same order ( see also the appendix of \cite{Khemchandani:2008rk}). The same happens when systems of three mesons are studied within the same chiral unitary approach \cite{MartinezTorres:2008gy}. This finding is most welcome because the results for the three body amplitudes do not depend on the unphysical off shell two body amplitudes. This cancellation obviously can not be seen in phenomenological approaches like 
\cite{Ikeda:2007nz,Shevchenko:2006xy,Shevchenko:2007zz,Ikeda:2008ub} since one cannot systematically associate the corresponding three body forces that would cancel the off shell contribution of the two body amplitudes. Thus, the results are linked to the off shell extrapolation of the amplitudes, and though one should expect cancellations with some three body forces, there is no way in those approaches to estimate the extend of these cancellations and the repercussion in the final binding energy. In view of it, awaiting hopefully a calculation in the line of   \cite{MartinezTorres:2007sr,Khemchandani:2008rk,MartinezTorres:2008gy}, which is not technically easy, we want to present a simplified approach, yet, realistic enough, which explicitly relies only on the on shell amplitudes of the interacting pairs and takes into account the energy dependence of the $\bar{K}N$ amplitude demanded by chiral dynamics. We find this possibility using the fixed center approximation to the Faddeev equations (FCA). As we shall discuss in the next section, the common findings of all the former approaches allow us to make the simplifying approximations that make the FCA (with some modifications)  an easy and reliable tool to face this problem. 

    The presence of the $\Lambda(1405)$ in the $\bar{K}N$ amplitude is one complicating factor because it is well known that all present chiral theories \cite{Kaiser:1995eg,angels,ollerulf,Lutz:2001yb,Oset:2001cn,Hyodo:2002pk,cola,Borasoy:2005ie,Oller:2006jw,Borasoy:2006sr} predict two poles close by in the 1405 MeV region. This means two states of the coupled channels $\pi \Sigma$, $\bar{K}N$ and other coupled channels. As found in \cite{cola} this had as a consequence that different amplitudes involving the  $\Lambda(1405)$ have different shapes, and as a consequence, the $\Lambda(1405)$ should appear with different shapes in different reactions. Since the pole with lower energy couples mostly to $\pi \Sigma$ and the one at higher mass mostly to $\bar{K}N$
\cite{cola}, it was expected that in reactions where the  $\Lambda(1405)$ production would be driven by the  $\bar{K}N$ the peak would appear around 1420 MeV, where all different chiral approaches predict this second pole. It did not take long to see this realized. Indeed, the study of the $K^- p \to \pi^0 	\pi^0 \Sigma^0$ reaction at $p_K = 514$ MeV/c - $750$ MeV/c  \cite{Prakhov:2004an} revealed the peak of the $\Lambda(1405)$ clearly at 1420 MeV. This was interpreted in \cite{Magas:2005vu} as being due to a dominant mechanism in which a pion is emitted from the initial nucleon, forcing the $\Lambda(1405)$ to be created from the $K^- p$ state. More recently, it was also found in \cite{Braun:1977wd} that a clear peak appeared for the $\Lambda(1405)$ in the $K^- d \to n \pi \Sigma$ reaction for kaons in flight. This is surprising in view that the $K^- p$ threshold is about 30 MeV above the $\Lambda(1405)$ mass. However, it was explained in  \cite{Jido:2009jf} that the peak was due to the rescattering of the kaon on a second nucleon, after it would lose some energy in the first collision against one nucleon and have the right kinematics to produce the $\Lambda(1405)$ in a second collision. The second surprise was that the peak of the resonance appears around 1420 MeV, not 1405 MeV, which is not surprising at all from the chiral dynamical perspective, since the resonance production is once again induced by the $\bar{K}N$ channel. A recent calculation \cite{Jido:2010rx} shows that the same features can already be observed for kaons produced at the DAFNE facility, and an experimental proposal is being prepared \cite{nevio}.

  Coming back to the three body works, the approach of \cite{Dote:2008in,Dote:2008hw} is fully consistent with these chiral dynamical 
features since it uses chiral theory as the underlying framework to construct the effective potential. This is not the case of the Faddeev 
approaches of \cite{Ikeda:2007nz,Shevchenko:2006xy,Shevchenko:2007zz,Ikeda:2008ub}, where to solve the AGS equations 
an energy independent kernel is used. In addition,
 providing just one $\Lambda(1405)$ resonance, the important dynamics of the two $\Lambda(1405)$ resonances is lost. Conscious about that, 
the authors of these works try to evaluate uncertainties by fitting their parameters to different masses of the $\Lambda(1405)$ from 1405 
MeV to 1420 MeV, and differences in the binding of about 10-20 MeV appear. The fits with a mass of the $\Lambda(1405)$ of 1420 MeV provide less
 binding, closer to the results of the variational approach of  \cite{Dote:2008in,Dote:2008hw}.

 An important step to conjugate the AGS equations
with the $\bar{K} N-\pi \Sigma$ dynamics of the chiral Lagrangians has been done, as mentioned, in \cite{Ikeda:2010tk}. Indeed, two poles
are found in qualitative agreement with other chiral approaches. One narrow pole around 1420 MeV, rather stable against changes of parameters,
is found in agreement with all findings of the chiral unitary approach. The second, wider pole, is found at very low energies 1335-1341 MeV, and 
more unstable with respect to changes of parameters. This agrees qualitatively with the findings of the chiral unitary approach, but the energy
is lower than in other approaches. One point to try to understand these difference is that in \cite{Ikeda:2010tk} the $\pi \Sigma$ mass 
distribution of the $K^- p\rightarrow\pi\pi\pi\Sigma$ in the Hemingway experiment \cite{Hemingway:1984pz} is adjusted assuming that it is proportional
to $|T_{\pi\Sigma,\pi\Sigma}|^{2}$. Yet, as shown in \cite{cola}, when one has two poles, the $T_{\pi\Sigma,\pi\Sigma}$
 and $T_{\bar{K} N,\pi\Sigma}$
amplitudes are rather different and the $\Lambda(1405)$ production processes proceed via the combination of the two amplitudes
$|T_{\pi\Sigma,\pi\Sigma}+ \beta ~T_{\bar{K} N,\pi\Sigma}|^{2}$ \cite{Hyodo:2003jw}. The fact that a separable potential with form factors 
is used also induces off shell effects with respect to the use of the on shell scattering amplitudes of the chiral unitary approach. All these 
facts certainly introduce uncertainties that revert in the deduced $T_{\pi\Sigma,\pi\Sigma}$, $T_{\bar{K} N,\pi\Sigma}$ amplitudes and the position
of the poles. These uncertainties should then translate into the position of the poles of the $\bar{K} NN-\pi \Sigma N$ system reported in 
\cite{Ikeda:2010tk}. In addition one would like to know the $T_{\bar{K} NN,\bar{K} NN}$ amplitude resulting from \cite{Ikeda:2010tk},
which was not provided in the paper and which would get contribution from the two poles of the three body system. This information would be useful 
to eventually compare with experiments, and in the case of the $\bar{K} N-\pi\Sigma$ system, the physical amplitudes do not have the shapes
of either of the poles alone.
  
   In the approach that we will use we shall follow the chiral unitary approach for the $\bar{K}N $ amplitudes, which provide the most 
important source of the binding of the three body system, according to the former studies. But there is also another different technical 
aspect of the present calculation with respect to the former ones. All previous approaches have concentrated on looking for the binding,
 searching for poles in the complex plane or looking for the energy that minimizes the expectation value of the Hamiltonian.  Here, inspired 
by the studies of   
\cite{MartinezTorres:2007sr,Khemchandani:2008rk,MartinezTorres:2008gy}, we shall look for bumps in the scattering matrices as a function of
 the energy of the three body system. Furthermore, the amplitudes obtained here could in principle be used as input for final state 
interaction when evaluating cross sections in reactions where eventually this  $\bar{K}N N$ state is formed.

The discussion done here has shown that, in spite of the much and good work done so far on the subject, uncertainties remain and further 
studies like the one done here should be welcome. Although there are certainly approximations in the FCA with respect to the Faddeev approach,
as there are also in the approximate three body scheme of the AGS equations, one can rely on some findings of these more elaborate approaches
to make the FCA results more solid, while it allows to overcome the sources of uncertainties mentioned in the other approaches and provide
a more intuitive and direct approach to the problem with the consequent transparency in the results obtained. Another of the novel things in the present 
work is that we show the existence of a less bound state in $S=1,~I=0$ of the two nucleons. This state was not considered in 
\cite{Shevchenko:2007zz,Dote:2008in,Dote:2008hw,Ikeda:2008ub,nogami,akayama,Ikeda:2010tk} although hints of the existence could be guessed from the large
negative real part of the $K^- d$ scattering length deduced in the work discussed in next section.

\section{Fixed Center formalism for the $\bar{K} (N N)$ interaction}

  The findings of the former works simplify our task. In this sense we should mention that in \cite{Dote:2008in} the nucleons are found to move slowly, having around 20 MeV of kinetic energy. It was also found there that the dependence on the type of NN interaction is rather weak. We rely upon the results of 
\cite{MartinezTorres:2007sr,Khemchandani:2008rk,MartinezTorres:2008gy} in the sense that only on shell two body amplitudes are needed. By this we mean the part of the analytical amplitudes obtained setting $q^2=m^2$ for the external particles, even if the particles are below threshold.

We also assume, like in the other works, that the two body interactions proceed in L=0. The effect of the p-wave interaction in the main interacting pair, the $\bar{K}N$ was found to be small in \cite{Dote:2008hw}, the main consequence being a moderate increase in the width, so we should accept the width that we find as a lower bound of  the actual width. 

  According to all the works, the main component of the wave function corresponds to having a $\bar{K} N $ in I=0 and hence the total isospin will be I=1/2. The total spin can be J=0,1, but the  $J^P=0^-$ state is the one found most attractive. We shall investigate both possibilities. 
  
   According to \cite{Shevchenko:2007zz} the effect of the $\Sigma N$ interaction is very small. We shall neglect it in our approach.
Yet we shall incorporate the $\pi\Sigma N $ as an explicit channel in the three body system, which in the $\bar{K}NN \rightarrow \bar{K}NN$
amplitude appears via the $\bar{K}N \rightarrow \pi\Sigma$ transition followed by $\pi N \rightarrow \pi N$ rescattering and 
$ \pi\Sigma \rightarrow \bar{K}N $ recombination.
   
   All this considered, the problem we shall solve is the interaction of a $\bar{K}$ with a $NN$ cluster.
 We shall evaluate the scattering matrix for this system as a function of the total energy of the  $\bar{K}NN$ system and look for peaks in $|T|^2$.
In a second step we will allow explicitly the intermediate  $\pi\Sigma$ state in the three body system.
We shall use in our study the FCA approximation to the $\bar{K}NN$ system. This approach has been used before in connection with the evaluation $K^{-}d$  
scattering length  \cite{Toker:1981zh,Chand:1962ec,Barrett:1999cw,Deloff:1999gc,Kamalov:2000iy} follow up of \cite{Kamalov:2000iy}
is done in \cite{Meissner:2006gx}. A discussion of these different approaches is done in \cite{Gal:2006cw}, where it is shown that Ref. \cite{Barrett:1999cw, Deloff:1999gc}
do not take into account explicitly the charge exchange $K^{-}p\rightarrow\bar{K}^{0}n$ reactions and antisymmetry of the nucleons, while 
it is explicitly done in Ref. \cite{Kamalov:2000iy}. In the present approach we also consider the antisymmetry and charge exchange in our scheme
although in a different technical way than in \cite{Kamalov:2000iy}, and assuming isospin symmetry for the $\bar{K}N$ interaction.

The FCA for $K^{-}d$ at threshold was found to be an acceptable approximation, within a few percent, to the more elaborate Faddeev equation
in \cite{Toker:1981zh,Gal:2006cw}. There it was also discussed that neglecting the $\pi\Sigma$ channel in the three body system is a good 
approximation, as far as $\bar{K}N$ interaction is evaluated in terms of the $\bar{K}N$ and $\pi\Sigma$ coupled channels. We shall be able to 
corroborate this here by explicitly introducing the $\pi\Sigma$ channel in the 3 body system.

Technically we follow closely the formalism of \cite{multirho}, where the FCA has been considered, using chiral amplitudes, in order to
 study theoretically the possibility of forming multi-$\rho$ states with large spin.

An interesting shared result of all the $K^{-}d$ calculation quoted above is a large and negative real part of the scattering length, of the order 
of $-1.40-1.80$ fm, which suggest the existence of a bound state in $ J=1$. Of course the imaginary part of the scattering length is equally
large, anticipating a broad state. We shall be able to make this more quantitative in the present approach. However, we shall also find that the most 
bound ${\bar K}NN $ system correspond to $J=0$, in agreement with other approaches.

The $NN$ interaction is of long range and very strong.
It binds the deuteron in spin $S=1$ and $I=0$, with $L=0$ to a very good approximation.
It nearly binds the $pp$ system also.
The binding of the $pp$ system is so close that a little help from an extra interaction, the strongly attractive ${\bar K}N$ interaction, is enough to also bind this system and we shall assume that this is the case here for $pp$ or in general for two nucleons in $S=0$, $I=1$, $L=0$.

Our strategy will be to assume as a starting point that the $NN$ cluster has a wave function like the one of the deuteron (we  omit the d-wave).
Latter on we will release this assumption and assume that the $NN$ system is further compressed in the ${\bar K}NN$ system.
Since in the FCA the input from the $NN$ system is the $NN$ form factor, taking into account an extra compression of the $NN$ systems is very easy by smoothly modifying the form factor to have a smaller radius.
As mentioned in the Introduction, we take information from previous studies, and in this sense it is interesting to recall that the calculations of \cite{galreport} for ${\bar K}$ bound in nuclei point out to a moderate compression of the nucleus due to the strong ${\bar K}N$ interaction.
However, in the case of two isolated nucleons the decrease in the radius can be far bigger that in nuclei, where nucleons are already
close to saturation density. The information on the $NN$ radius in the ${\bar K}NN$ molecule we get from \cite{Dote:2008hw}, where the $NN$
interaction is taken into account including short range repulsion that precludes unreasonable compression. Yet, the r.m.s radius found is of 
the order of 2.2 fm, slightly above one half the value of the deuteron r.m.s radius of 3.98 fm ($NN$ distance).

\subsection{Calculation of the three body amplitude}
The formal derivation follows the steps of \cite{multirho}.
We assume pure $L=0$ interaction for all the ${\bar K}N$ and $NN$ pairs, hence we have $S=J=0$, $I_{NN}=1$ or $S=J=1$, $I_{NN}=0$ for the two nucleons and we shall evaluate the interaction in both cases.
Since $L=0$ also for ${\bar K}N$, $J$ will be the spin of the total ${\bar K}NN$ system.
In the case of $I_{NN}=0$ for $NN$, the total isospin will be $I=1/2$.
In the case of $I_{NN}=1$ we shall also couple the total isospin $I$ to $1/2$ to give room to the $\Lambda(1405)N$ configuration which is where the attraction concentrates according to former studies.

\begin{figure}[!t]
\begin{center}
\includegraphics[width=0.75\textwidth]{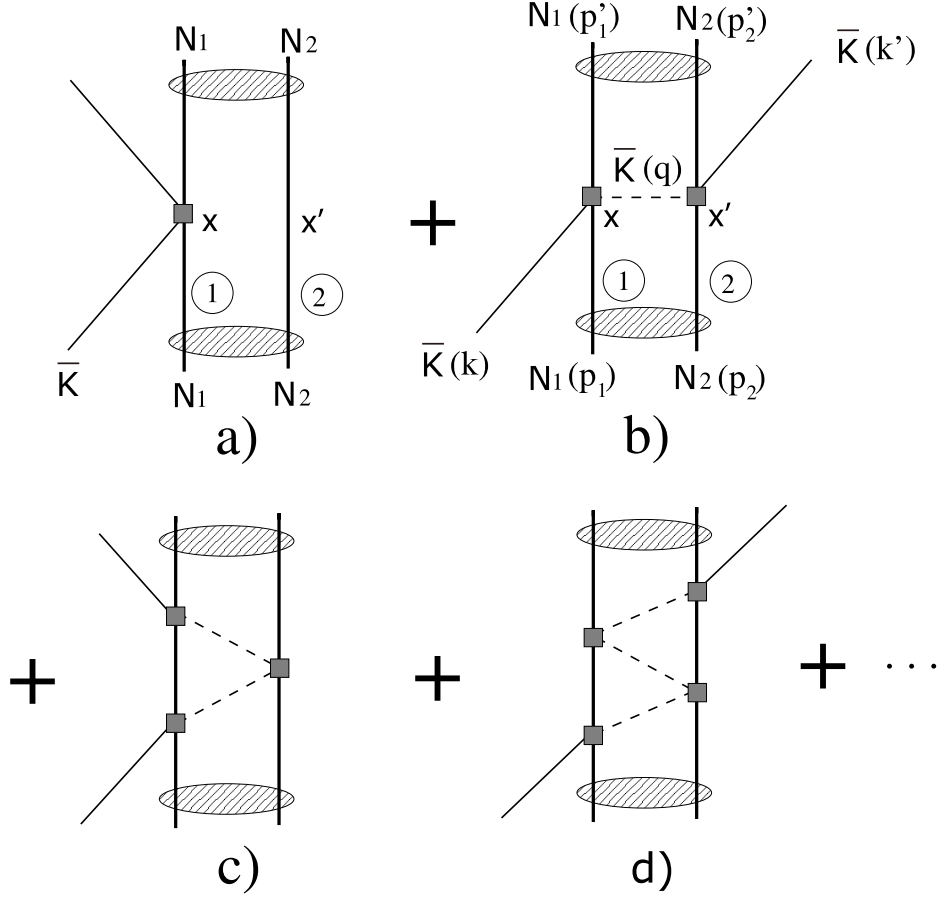}
\caption{Mechanism for the ${\bar K}NN$ interaction including multiple scattering of the ${\bar K}$ with the nucleons.
The equivalent diagrams where the ${\bar K}$ interacts first with the second nucleon should be added.}
\label{fig:1}
\end{center}
\end{figure}

The mechanism implicit in the FCA are depicted in Fig.~\ref{fig:1}. The first thing we need is to evaluate the interaction of a ${\bar K}$ with either of the two nucleons in each of the two configurations that we will consider: $S=0$, $I_{NN}=1$, $I=1/2$ and $S=1$, $I_{NN}=0$, $I=1/2$.
The ${\bar K}N$ scattering matrix is obtained as follows. First we write the wave function for the states.

\begin{align}
|\bar{K}(NN)_{I_{NN}=0},I=1/2,I_{z}=1/2>=|1/2>\dfrac{1}{\sqrt{2}}(|1/2,-1/2>-|-1/2,1/2>)\label{Eq:amel}%
\end{align}
with the nomenclature $|I_{z}>$ for the states of $\bar{K}$ and $|I_{z_{1}},I_{z_{2}}>$ for the $NN$ system. We want to evaluate
\begin{align}
<\bar{K}(NN)_{I_{NN}=0},I=1/2,I_{z}=1/2|t_{31}|\bar{K}(NN)_{I_{NN}=0},I=1/2,I_{z}=1/2>\label{Eq:amel2}%
\end{align}
and the same one for $t_{32}$, where $t_{31},~t_{32}$ stand for the scattering matrix of the ${\bar K}$ with nucleons
1 and 2, respectively.

In order to use the isospin scattering matrices we write the state of Eqs.~(\ref{Eq:amel}) in terms of good isospin states of
${\bar K}N$ as 
\begin{align}
\dfrac{1}{\sqrt{2}}|1/2>(|1/2,-1/2>-|-1/2,1/2>)&= \dfrac{1}{\sqrt{2}}(|1,1>|-1/2>-\dfrac{1}{\sqrt{2}}|1,0>|1/2>
\nonumber\\&-\dfrac{1}{\sqrt{2}}|0,0>|1/2>)\label{Eq:amel3}%
\end{align}
where in the second member of the equation we use the nomenclature $|{\bar K}N,I,I_{z}>|I_{N_{2}z}>$ and then
we find immediately that 
\begin{align}
I_{NN}=0~:~t_{31}=\dfrac{3}{4}t_{{\bar K}N}^{I=1}+\dfrac{1}{4}t_{{\bar K}N}^{I=0}\label{Eq:amel4}%
\end{align}
Similarly, combining the ${\bar K}$ isospin with that of the second nucleons we find
\begin{align}
I_{NN}=0~:~t_{32}=\dfrac{3}{4}t_{{\bar K}N}^{I=1}+\dfrac{1}{4}t_{{\bar K}N}^{I=0}\label{Eq:amel5}%
\end{align}

Similarly we evaluate $t_{31},~t_{32}$ for the case of $I_{NN}=1$

\begin{align}
|\bar{K}(NN)_{I_{NN}=1},I=1/2,I_{z}=1/2>=-\sqrt{\dfrac{2}{3}}|1/2,-1/2>|1,1>+\sqrt{\dfrac{1}{3}}|1/2,1/2>|1,0>
\label{Eq:amel6}%
\end{align}
with nomenclature $|I_{{\bar K}},I_{{\bar K},z}>|I_{NN},I_{NN,z}>$
\begin{align}
|\bar{K}(NN)_{I_{NN}=1},I=1/2,I_{z}=1/2>&=-\sqrt{\dfrac{2}{3}}|-1/2>|1/2,1/2>\nonumber\\&
+\dfrac{1}{\sqrt{3}}|1/2>(\dfrac{1}{\sqrt{2}}|1/2,-1/2>+\dfrac{1}{\sqrt{2}}|-1/2,1/2>)
\label{Eq:amel7}%
\end{align}
where now we are using the nomenclature $|{\bar K},I_{z}>|I_{N_{1}z},I_{N_{2}z}>$.
Combining $\bar{K}N_{1}$ to have good isospin states we obtain 
\begin{align}
|\bar{K}(NN)_{I_{NN}=1},I=1/2,I_{z}=1/2>&=-\sqrt{\dfrac{2}{3}}\dfrac{1}{\sqrt{2}}|1,0>|1/2>+\sqrt{\dfrac{2}{3}}\dfrac{1}{\sqrt{2}}|0,0>|1/2>
\nonumber\\&+\dfrac{1}{\sqrt{6}}|1,1>|-1/2>+\dfrac{1}{\sqrt{6}}\dfrac{1}{\sqrt{2}}|1,0>|1/2>
\nonumber\\&+\dfrac{1}{\sqrt{6}}\dfrac{1}{\sqrt{2}}|0,0>|1/2>
\label{Eq:amel8}%
\end{align}
with nomenclature $ |\bar{K}N,I,I_{z}>|I_{N_{2},z}> $, and then we find immediately that 
\begin{align}
I_{NN}=1~:~t_{31}=\dfrac{1}{4}t_{{\bar K}N}^{I=1}+\dfrac{3}{4}t_{{\bar K}N}^{I=0}\label{Eq:amel9}%
\end{align}
and similarly
\begin{align}
I_{NN}=1~:~t_{32}=\dfrac{1}{4}t_{{\bar K}N}^{I=1}+\dfrac{3}{4}t_{{\bar K}N}^{I=0}.\label{Eq:amel10}%
\end{align}

The $S$ matrix for the single scattering of ${\bar K}N_{1,2}$ (Fig.~\ref{fig:1} a)) is given by
\begin{eqnarray}
S^{(1,1)}&=&-it_{1} \frac{1}{{\cal V}^2}
\frac{1}{\sqrt{2\omega_{\bar K}}}
\frac{1}{\sqrt{2\omega'_{\bar K}}}
\sqrt{\frac{2M}{2E_1}}
\sqrt{\frac{2M}{2E'_1}}\nonumber\\
&&\times(2\pi)^4\,\delta(k+k_{NN}-k'-k'_{NN}),\\
S^{(1,2)}&=&-it_{2} \frac{1}{{\cal V}^2}
\frac{1}{\sqrt{2\omega_{\bar K}}}
\frac{1}{\sqrt{2\omega'_{\bar K}}}
\sqrt{\frac{2M}{2E_2}}
\sqrt{\frac{2M}{2E'_2}}\nonumber\\
&&\times(2\pi)^4\,\delta(k+k_{NN}-k'-k'_{NN}).
\end{eqnarray}
while that for the double scattering, Fig.~\ref{fig:1} b), it is given by
\begin{eqnarray}
S^{(2)}&=&-i(2\pi)^4 \delta(k+k_{NN}-k'-k'_{NN})\frac{1}{{\cal V}^2}
\frac{1}{\sqrt{2\omega_{\bar K}}} 
\frac{1}{\sqrt{2\omega'_{\bar K}}}
\sqrt{\frac{2M}{2E_1}}
\sqrt{\frac{2M}{2E'_1}}
\sqrt{\frac{2M}{2E_2}}
\sqrt{\frac{2M}{2E'_2}}\nonumber\\
&&\times\int \frac{d^3q}{(2\pi)^3} 
F_{NN}(q)
\frac{1}{{q^0}^2-\vec{q}\,^2-m_{\bar K}^2+i\epsilon} t_{1} t_{2}.
\end{eqnarray}
where $F_{NN}(q)$ is the form factor of the $NN$ system.

On the other hand, if we consider the scattering of ${\bar K}$ with the $NN$ system the $S$ matrix is given by
\begin{eqnarray}
S&=&-i T(2\pi)^4 \delta(k+k_{NN}-k'-k'_{NN})\frac{1}{{\cal V}^2}\nonumber\\
&&\times\frac{1}{\sqrt{2 \omega_{\bar K}}} 
\frac{1}{\sqrt{2 \omega'_{\bar K}}}
\sqrt{\frac{2M_{NN}}{2E_{NN}}}
\sqrt{\frac{2M_{NN}}{2E'_{NN}}}.
\label{Eq:ss}
\end{eqnarray}

As we can see, the field normalization factors that appear in the amplitude of the different terms, for which we follow the Mandl and Shaw convention \cite{mandl}, are different.
However, if we approximate $\omega_{\bar K}=m_{\bar K}$, $\sqrt{M_N/E_N}=\sqrt{2M_{NN}/2E_{NN}}=1$, then all the normalization factors are the same.
Then we can write $T$ in terms of two partitions, $T_1$,~$T_2$, summing all the terms which begin with the ${\bar K}$ interaction with the nucleon 1 or nucleon 2, respectively.
Then we find
\begin{equation}
T=T_1+T_2
\end{equation}
\begin{eqnarray}
T_1&=&t_1+t_1G_0T_2
\label{Eq:a}\\
T_2&=&t_2+t_2G_0T_1\nonumber
\end{eqnarray}
where 
\begin{equation}
G_0=\int\frac{d^3q}{(2\pi)^3}F_{NN}(q)\frac{1}{{q^0}^2-\vec{q}\,^2-m_{\bar K}^2+i\epsilon}.
\label{Eq:b}
\end{equation}
One can easily solve Eqs.~(\ref{Eq:a}) and we obtain
\begin{equation}
T=T_1+T_2=\frac{t_1+t_2+2t_1t_2G_0}{1-t_1t_2G_0^2}
\label{Eq:c}
\end{equation}
which, taking into account that $t_1=t_2$  can be written as
 \begin{equation}
T=\frac{2t_1}{1-t_1G_0}.
\label{Eq:cla}
\end{equation}

In principle one must also consider the form factor in the single scattering \cite{multirho}, but as done in \cite{multirho}, taking into account that one has small momenta of ${\bar K}$ in the bound ${\bar K}N$ system, one can just take it equal 1.

The variable $q^0$ in Eq.~(\ref{Eq:b}) is the energy carried by the ${\bar K}$, which is given by 
\begin{equation}
q^0=\frac{s+m_{\bar K}^2-(2M_N)^2}{2\sqrt{s}}
\end{equation}
with $\sqrt{s}$ the rest energy of the ${\bar K}NN$ system.
We also need the argument of the $t_1$ and $t_2$ function, $s_1$, given by
\begin{equation}
s_1=m^2_{\bar K}+m_N^2+\frac{1}{2}(s-m_{\bar K}^2-4m_N^2)~~.
\end{equation}

For the form factor we take, as a starting point, the one of the deuteron which is given by 
\begin{align}
&  F(q)= \int^{\infty}_{0} d^{3}p ~\sum_{j=1}^{11} \frac{C_{j}}{\vec{p}^{2}+m_{j}^{2}}\sum_{i=1}^{11}\frac{C_{i}}{(\vec{p}-\vec{q})^{2}+m_{i}^{2}}, \label{formfac}%
\end{align}  
where the $C_{i}$ coefficients are given in \cite{Machleidt:2001dd}. The form factor is normalized to 1 at $\vec{q}=0$ by 
dividing $F(q)$ by the expression of Eq.~(\ref{formfac}) at $\vec{q}=0$. In further steps this form factor will be changed to accomodate the reduced size of 
the two $N$ system found in \cite{Dote:2008hw}. This is done rescaling the masses $m_i$ appearing in Eq.~(\ref{formfac}), demanding that the radius
be the one of Ref. \cite{Dote:2008hw}.

\section{Results}

\begin{figure}[ptbh]
\begin{center}
\includegraphics[width=1\textwidth]{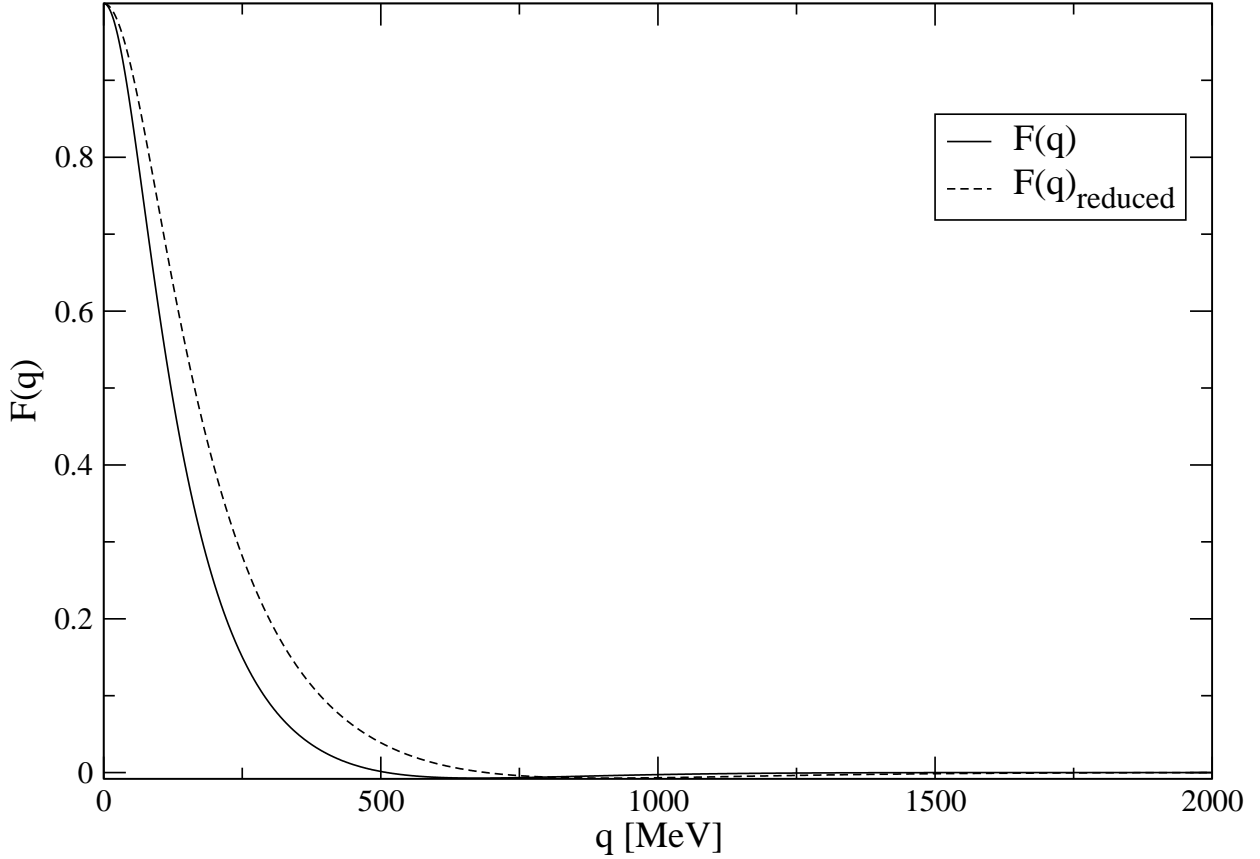}
\caption{Form factor of the deuteron, and the one corresponding to an $NN$ system with a reduced 
radius from Ref. \cite{Dote:2008hw}.}
\label{fig:fff}
\end{center}
\end{figure}

\begin{figure}[ptbh]
\begin{center}
\includegraphics[width=1\textwidth]{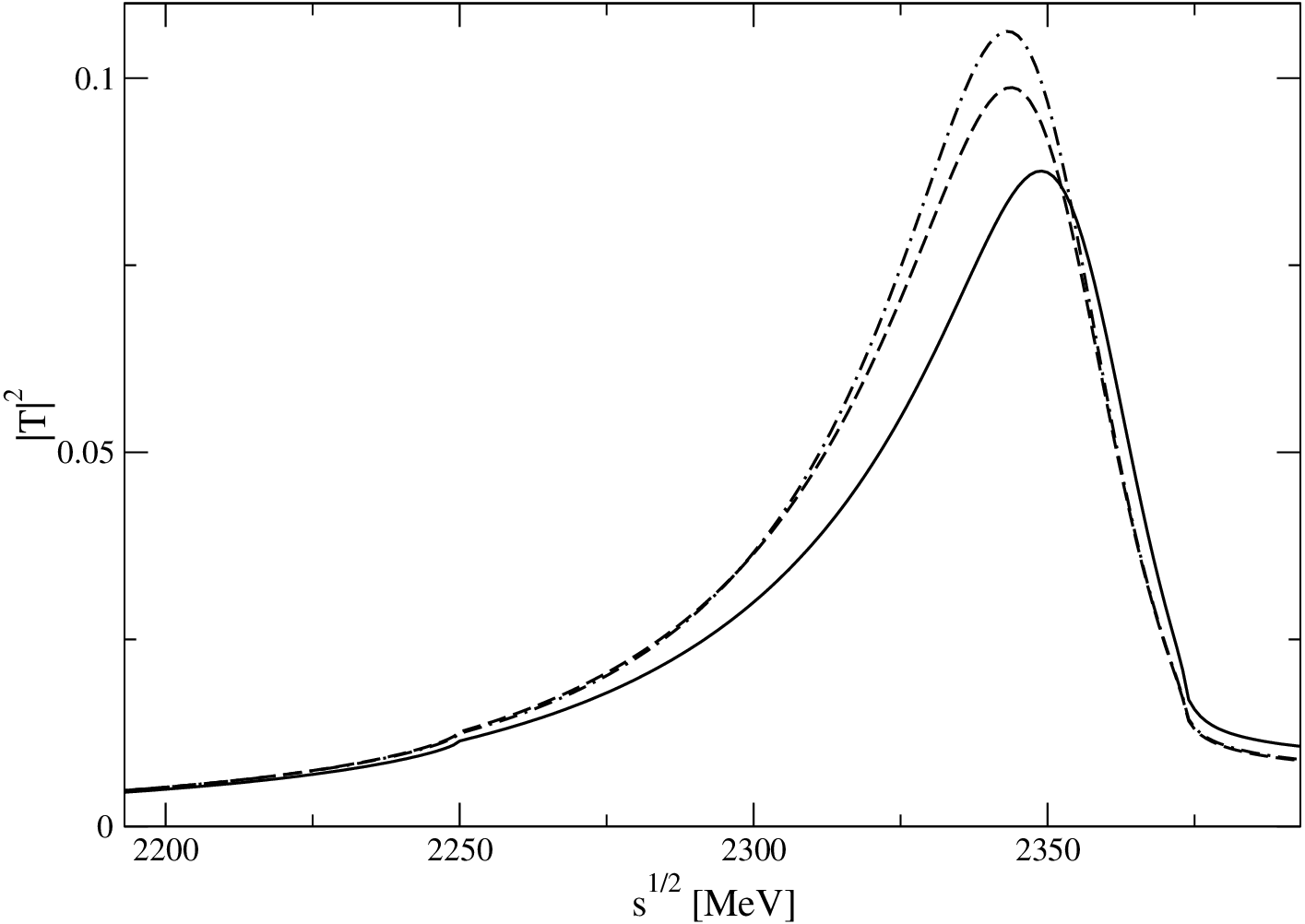}
\caption{Modulus squared of the three-body scattering amplitude for triplet (S=1). The solid line indicates the $\bar{K}NN$ system in normal size 
and the dashed line indicates the same system with reduced $NN$ radius Ref. \cite{Dote:2008hw}. The dot-dashed line indicates the result of the including
 the $\pi \Sigma$ channel with the reduced radius. }
\label{fig:tmats1}
\end{center}
\end{figure}

In Fig.~\ref{fig:fff} we plot the deuteron form factor $F(q)$ normalized to 1 at $q=0$, as well as the one corresponding to the reduced
$NN$ radius, as described in the former section.  
In Fig.~\ref{fig:tmats1} we show the results for $|T|^{2}$ of  Eq.~(\ref{Eq:cla}), as a function of the total energy of the $\bar{K}NN$ system
for the case of $S=1$. The results obtained using  $F(q)$ from the deuteron show a clear peak at $\sqrt{s}=2350$ MeV, about $22$ MeV
below the threshold of $\bar{K}NN$. The width of the distribution is about $50$ MeV. It is interesting to see what happens with
the decreased $NN$ radius. As one can see, the effects are clearly visible, the binding is increased
by about five MeV and the width is practically unchanged.

\begin{figure}[ptbh]
\begin{center}
\includegraphics[width=1\textwidth]{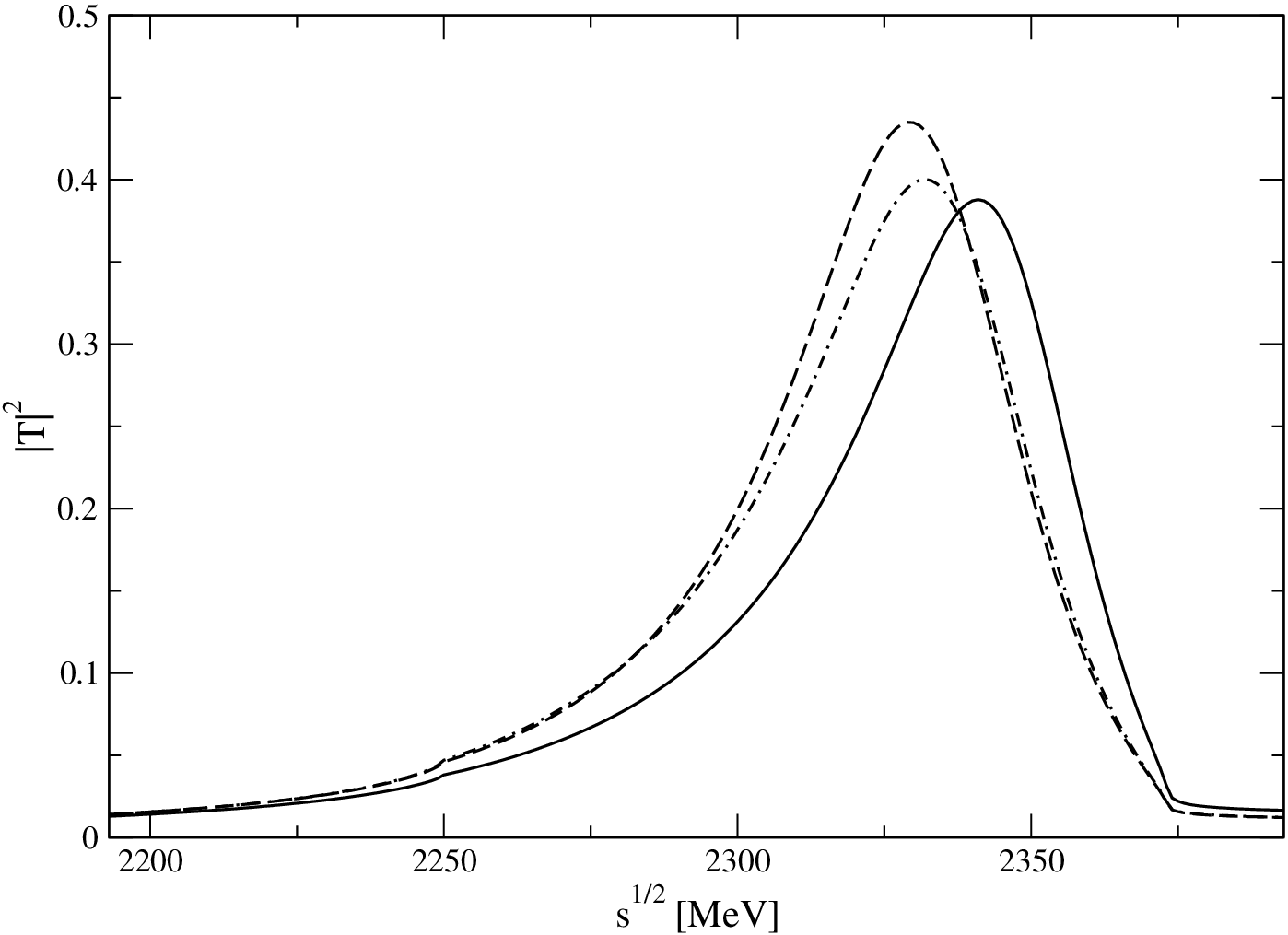}
\caption{Modulus squared of the three-body scattering amplitude for singlet (S=0).  The solid line indicates the $\bar{K}NN$ system in normal size 
and the dashed line indicates the same system with reduced $NN$ radius Ref. \cite{Dote:2008hw}. The dot-dashed line indicates the result of the including
the  $\pi \Sigma$ channel with the reduced radius.}
\label{fig:s0}
\end{center}
\end{figure}

  More interesting is to see what happens for $S=0$. This is shown in Fig.~\ref{fig:s0}. We find here that $|T|^{2}$ also shows 
a peak, but more bound than for $S=1$. The position of the peak indicates a binding of the  $\bar{K}NN$ system by $30$ MeV. 
The width is of about $50$ MeV. When we take the case of the reduced $NN$ radius the peak is displaced to about $40$ MeV binding and 
the width is still around $50$ MeV.  

  The results obtained could be interpreted as giving support to two independent states since they have different total angular
 momentum, but some considerations are in order. The first important thing is to observe  the different strength of the
 two distributions. Indeed 
$|T|^{2}$ is $4.5$ times bigger in the $S=0$ case than in $S=1$. Second, in the case of $S=0$ the system is more bound.
 The two independent states come in our approach because we have assumed exact $L=0$ contribution.
Should we allow for some $L=1$ contribution in the $\bar{K}N$ interaction, we can have a contribution
to total spin $J=0$ from a spin $S=1$ of the two nucleons and then the different spins of the nucleons would mix in the real state.
But this mixing should be small in view of the weak effects of the $\bar{K}N$ p-wave reported in \cite{Dote:2008hw}.

 It is clear from our results that the state with $J=0$ will correspond to the most bound $\bar{K}NN$ system and that it will contain 
a small admixture of $S=1(I=0)$ of the two nucleons. This is the same conclusion reached in \cite{Dote:2008in,Dote:2008hw}.

The approach followed here shows with clarity why the  $S=0,~ I_{NN}=1$ configuration is preferred. Indeed, as one can see 
in Eqs.~(\ref{Eq:amel4}),(\ref{Eq:amel5}),(\ref{Eq:amel9}),(\ref{Eq:amel10}) in the case of $S=1,~ I_{NN}=0$ the $t_{1,2}$ amplitudes have a weight of $3/4$ for $I_{\bar{K}N}=1$ and $1/4$
for  $I_{\bar{K}N}=0$, while in the case of $S=0,~ I_{NN}=1$ the weights are opposite, with $1/4$ for $I_{\bar{K}N}=1$ and $3/4$
for $I_{\bar{K}N}=0$. Since the strong attraction appears in $I_{\bar{K}N}=0$, where the two $\varLambda(1405)$ states are generated,
it becomes clear that one gets more attraction and more strength in $|T|^{2}$ for the case of $S=0,~ I_{NN}=1$ than for $S=1,~ I_{NN}=0$.

On the other hand, the exercise made here also tells that, while the $S=1,~ I_{NN}=0,~L_{\bar{K}N}=0$ configuration does not correspond to a
pure state and does not lead to the most bound component of the $\bar{K}NN$ system, it is still a configuration with a sizeable strength,
although reasonably smaller than the one with $S=0,~ I_{NN}=1,~L_{\bar{K}N}=0$. In a reaction producing $\bar{K}NN$ (bound) in the final 
state, these configuration would all contribute to the cross section, resulting in a peak with larger width. Note that in such reactions
one would have to look for decay products of $\bar{K}NN$ to identify the peak that we produce. A possible channel would be $\varLambda N$,
 but one should not expect to see such a clean peak as we have produced, since the production amplitude with $\varLambda N$ in the final states 
will also have contribution from  ``uncorrelated''  $\varLambda p$ production which might have a large strength compared to the contribution 
of $\varLambda p$ production mediated by the doorway mechanism of $\bar{K}NN$ (bound) production. An answer to this issue can only 
be provided by detailed calculations for each particular reaction. In such a case, the explicit evaluation of the $\bar{K}NN$ scattering
amplitude that we have obtained would be of much use.

The width of the $\bar{K}NN$ state of about $50$ MeV that we have obtained is in line with that obtained in most approaches. Some of them
provide even larger widths  \cite{Shevchenko:2006xy,Shevchenko:2007zz,Dote:2008in,Dote:2008hw}. Note that although our  $\bar{K}N$ amplitudes contain the transition
$\bar{K}N\rightarrow \pi \varSigma $  in intermediate states, and hence our approach contains $\pi \varSigma N$ intermediate states,
we do not have diagrams in our approach, so far, in which the $\pi$ or the $\varSigma$ interact with the nucleon,  though former work have
indicated these effects to be small, as mentioned in the introduction. We come back to this issue in the next section. Common to all the theoretical approaches is the lack of 
$\bar{K}$ absorption by two nucleons. In our approach, this is also not considered, and this would also make the width larger. Estimates
done in \cite{Dote:2008in,Dote:2008hw} point to one increase of extra $25$ MeV in the width from this source, leading to widths considerably
larger than the binding, which added to other difficulties pointed out above, do not help when trying to observe experimentally
this interesting state.

\section{Explicit considiration of the $\pi \Sigma N $ channnel in the three body system}

\begin{figure}[!t]
\begin{center}
\includegraphics[width=0.35\textwidth]{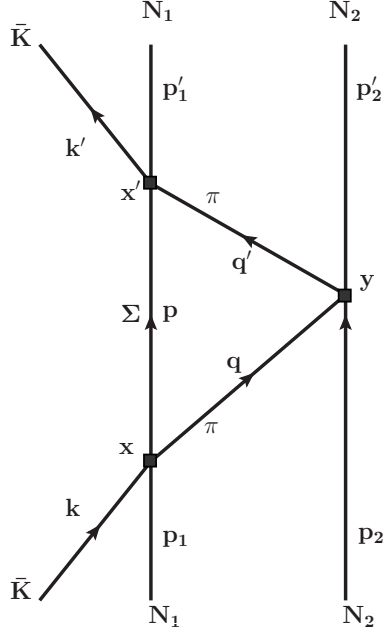}
\caption{Feynman diagram  for the $\pi \Sigma N$  channel.
The equivalent diagrams where the ${\bar K}$ interacts first with the second nucleon should be added.}
\label{fig:10}
\end{center}
\end{figure}

As we have explained, the $\pi \Sigma$ channel and other coupled channels are explicitly taken into account in the consideration of the 
$\bar{K}N$ amplitude which we have used in the FCA approach. This means that we account for the $\bar{K}N \rightarrow  \pi \Sigma$ transition,
and an intermediate  $\pi \Sigma N $ channel, but this  $\pi \Sigma $ state is again reconverted to $\bar{K}N$ leaving the other $N$
as a spectator. This is accounted in the multiple scattering in the  $\bar{K}N \rightarrow \bar{K}N $ scattering matrix on one nucleon. However, we do not consider the possibility
that one has $\bar{K}N \rightarrow  \pi \Sigma$ and the $\pi$ rescatters with the second nucleon. If we want to have a final $\bar{K}N N$
system again, the $\pi$ has to scatter later with the $\Sigma$ to produce $\bar{K}N$. One may consider multiple scatterings of the $\pi$ 
with the nucleons, but given the smallness of the $\pi N$  amplitude compared to the $\bar{K}N$, any diagram beyond the one having one rescattering
of the pion with the nucleon will be negligible. Then we must consider the diagram  of
Fig. ~\ref{fig:10} (for $\bar{K}N$ scattering on nucleon 1). 

The S matrix for this process is given by 

\begin{eqnarray}
S^{(\pi \Sigma)}&=& \int d^{4}x \int d^{4}x' \int d^{4}y \frac{1}{\sqrt{2\omega_{\bar K  }\cal V }} 
\frac{1}{\sqrt{2\omega_{\bar K'}\cal V }} e^{-ikx} e^{ik'x'} \varphi_{1}(x) \varphi_{2}(y) \varphi^{*}_{1}(x') 
\varphi^{*}_{2}(y)\nonumber\\&&
\int  \frac{d^{4}p}{(2\pi)^{4}}  e^{-ip (x'-x)} \frac{2 M_{\Sigma}}{2 E_{\Sigma}(\vec{p})}
 \frac{1}{p^{0}-E_{\Sigma}(\vec{p})+i \epsilon}
 \nonumber\\&& \int  \frac{d^{4}q}{(2\pi)^{4}}   e^{-iq (y-x)} \frac{1}{q^{2}-m_{\pi}^{2}+i \epsilon}
\nonumber\\&& \int \frac{d^{4}q'}{(2\pi)^{4}} e^{-iq' (x'-y)} \frac{1}{q'^{2}-m_{\pi}^{2}+i \epsilon}
t_{\bar{K}N,\pi \Sigma} ~t_{\pi N,\pi N}~ t_{\pi \Sigma,\bar{K}N}.
\end{eqnarray}
We use the variables 

  \begin{eqnarray}
\vec{R}=\frac{1}{2} (\vec{x}+\vec{y}) ,~~\vec{R'}=\frac{1}{2} (\vec{x'}+\vec{y}),~~\vec{r}=(\vec{y}-\vec{x}),~~\vec{r'}=(\vec{x'}-\vec{y})
,~~\vec{R'}=\vec{R}+\frac{1}{2} (\vec{r}+\vec{r'})
\end{eqnarray}
and perform explicitly the $x^{0},x'^{0},y^{0},q^{0},q'^{0},p^{0}$  integrations following similar steps as in Section II A.

We also write 
\begin{eqnarray}
 &&\varphi_{1}(\vec{x}) \varphi_{2}(\vec{y})= \frac{1}{\sqrt{\cal V} }  e^{i \vec{k}_{NN} \vec{R}} \varphi(\vec{r})\nonumber\\&&
\varphi^{*}_{1}(\vec{x'}) \varphi^{*}_{2}((\vec{y})= \frac{1}{\sqrt{\cal V} }  e^{-i \vec{k'}_{NN} \vec{R'}} \varphi^{*}(\vec{r'})
\end{eqnarray}
where $\varphi(\vec{r})$ is the wave function of the $NN$ system, for which we take the same expression as it was used to evaluate the form factor.
In momentum space 

\begin{eqnarray}
 \tilde{\varphi}(\vec{q})=\int d^{3}\vec{r} e^{i\vec{q} \vec{r}}  \varphi(\vec{r})=\frac{1}{N} ~\sum_{j} \frac{C_{j}}{\vec{q}^{2}+m_{j}^{2}}
\end{eqnarray}
 with 
\begin{eqnarray}
 N^{2}=\frac{1}{(2\pi)^{3}}\int d^{3}\vec{p} ~\bigg(\sum_{j} \frac{C_{j}}{\vec{p}^{2}+m_{j}^{2}}\bigg)^{2}.
\end{eqnarray}

Then we find 

\begin{eqnarray}
S^{(\pi \Sigma)}&=& i \frac{1}{\sqrt{2\omega_{\bar K  }\cal V }} \frac{1}{\sqrt{2\omega_{\bar K'}\cal V }}
\int  \frac{d^{3}\vec{q}}{(2\pi)^{3}} \int  \frac{d^{3} \vec{q'}}{(2\pi)^{3}} \int  \frac{d^{3} \vec{p}}{(2\pi)^{3}}
\frac{M_{\Sigma}}{E_{\Sigma}(\vec{p})} \tilde{\varphi}(\vec{p}+\vec{q}) \tilde{\varphi}(\vec{p}+\vec{q'})  \nonumber\\&&
(2\pi)^4 \delta(k+k_{NN}-k'-k'_{NN}) t_{\bar{K}N,\pi \Sigma} ~t_{\pi N,\pi N}~ t_{\pi \Sigma,\bar{K}N}
\nonumber\\&&
 \frac{1}{2} \frac{1}{\omega(\vec{q})+\omega (\vec{q'})} \frac{1}{\omega(\vec{q})~\omega (\vec{q'})} 
 \frac{1}{\sqrt{s_{1}}-\omega(\vec{q})-E(\vec{p})+i\epsilon} \nonumber\\&&
\frac{1}{\sqrt{s_{1}}-\omega (\vec{q'})-E(\vec{p})+i\epsilon} (\sqrt{s_{1}}-E(\vec{p})-\omega(\vec{q})-\omega (\vec{q'})).
\end{eqnarray}
One can also perform the angular integrations in $\vec{q},~\vec{q'}$ explicitly and considering the relationship of $S$ to $T$ from
Eq.~(\ref{Eq:ss}) we find 

\begin{eqnarray}
T^{(\pi \Sigma)}&=& - \frac{1}{N^{2} } \frac{1}{(2 \pi)^{3} }\int p^{2} d p  \frac{M_{\Sigma}}{E_{\Sigma}(\vec{p})} \int  q~dq \int  q'~dq'
\sum_{j} \frac{C_{j}}{2 p} \ln\frac{p^{2}+q^{2}+2 p q+m_{j}^{2}}{p^{2}+
q^{2}-2pq+m_{j}^{2}}\nonumber\\&&
\sum_{i} \frac{C_{i}}{2p} \ln\frac{p^{2}+q'^{2}+2pq'+m_{i}^{2}}{p^{2}+q'^{2}-2p
q'+m_{i}^{2}}~
t_{\bar{K}N,\pi \Sigma} ~t_{\pi N,\pi N}~ t_{\pi \Sigma,\bar{K}N}
\nonumber\\&&
\frac{1}{\omega(\vec{q})+\omega (\vec{q'})} \frac{1}{\omega(\vec{q})~\omega (\vec{q'})} 
 \frac{1}{\sqrt{s_{1}}-\omega(\vec{q})-E(\vec{p})+i\epsilon} \nonumber\\&&
\frac{1}{\sqrt{s_{1}}-\omega (\vec{q'})-E(\vec{p})+i\epsilon} (\sqrt{s_{1}}-E(\vec{p})-\omega(\vec{q})-\omega(\vec{q'})).
  \label{Eq:tmatrixpisigma}
\end{eqnarray}

Using the same arguments as in Section II A one can show that the $t_{\bar{K}N,\pi \Sigma}$ amplitude that appears in
 Eq.~(\ref{Eq:tmatrixpisigma}) is given by 
\begin{align}
t_{\bar{K}N,\pi \Sigma}=\dfrac{1}{4}t_{\bar{K}N,\pi \Sigma}^{I=1}+\dfrac{3}{4}t_{\bar{K}N,\pi \Sigma}^{I=0}~;~S=0
\end{align}

\begin{align}
t_{\bar{K}N,\pi \Sigma}=\dfrac{3}{4}t_{\bar{K}N,\pi \Sigma}^{I=1}+\dfrac{1}{4}t_{\bar{K}N,\pi \Sigma}^{I=0}~;~S=1
\end{align}

\begin{align}
t_{\pi N,\pi N}=(\frac{7}{12}-\frac{1}{\sqrt{6}})t_{\pi N,\pi N}^{I=\frac{3}{2}}+
(\frac{5}{12}+\frac{1}{\sqrt{6}})t_{\pi N,\pi N}^{I=\frac{1}{2}}~;~S=0
\end{align}

\begin{align}
t_{\pi N,\pi N}=(\frac{11}{12}+\frac{2}{3}\frac{1}{\sqrt{3}}+\frac{2}{3}\frac{1}{\sqrt{6}}-\frac{1}{6}\sqrt{2})t_{\pi N,\pi N}^{I=\frac{3}{2}}+
(\frac{7}{12}-\frac{2}{3}\frac{1}{\sqrt{3}}-\frac{2}{3}\frac{1}{\sqrt{6}}+\frac{1}{6}\sqrt{2})t_{\pi N,\pi N}^{I=\frac{1}{2}}~;~S=1.
\end{align}
We get $t_{\pi N, \pi N}$ from the work of \cite{Inoue:2001ip} \footnote{ We thank A. Ramos for providing us with the numerical amplitudes
of the model of \cite{Inoue:2001ip}.}.
We find that $T^{(\pi \Sigma)}$ is about a factor 20 smaller than $t_{\bar{K}N,\bar{K}N}$. Once we have $T^{(\pi \Sigma)}$
we add it coherently to $t_{\bar{K}N,\bar{K}N}~ (t_{1})$ in  Eq.~(\ref{Eq:cla}) and recalculate the total $T$ matrix. This will also contain
rescattering effects. 

The result of the including the $\pi \Sigma$ channel can be seen in Figs.~\ref{fig:tmats1}, \ref{fig:s0}  for $S=1,S=0$.
We can see that the effects are moderate. For $S=0$ we find a small shift of the peak of about 2-3 MeV to smaller binding energies and a small broadening
of the width by about 2-3 MeV. The small shift of the peak to lower energies does not contradict the findings of 
\cite{Shevchenko:2007zz} where the comparison is made with Faddeev equations using $\bar{K}N$ and $\pi \Sigma$ channels or the same Faddeev
equations using only $\bar{K}N$, both in the two body scattering matrices and in the three body Faddeev equations. Here the 
$\pi \Sigma,\bar{K}N$ channels have been considered always in the two body $\bar{K}N$ amplitudes. This result answers a question
raised in the introduction, and the about 11 MeV shift of the binding energy seen in  \cite{Shevchenko:2007zz} should be mostly attributed to the consideration of the 
$\pi \Sigma$ channel in the two body scattering matrices rather than its explicit consideration in the extra diagrams of the Faddeev series that we have evaluated
explicitly here.

For the case of $S=1$ we can see the result in Fig. ~\ref{fig:tmats1}. The effects of the $T^{(\pi \Sigma)}$ amplitude are also small and here 
we do not see any visible change in the peak position or the width, only a small increase in the $|T|^{2}$.

\section{Conclusions} 

We performed calculations for the scattering amplitudes of the $\bar{K}NN$ system using the FCA of 
Faddeev equations and considering the scattering of the light $\bar{K}$ against the heavier NN cluster. 
We assume that a NN cluster is made, since in $S=1$, $I_{NN}=0$ the $NN$ system is bound, and the  $S=0$, 
$I_{NN}=1$, which is nearly bound in free space, gets the small push needed to bind from the strong attractive
$\bar{K}N$ interaction. 
We reduced the ``frozen'' condition for the $NN$ system by allowing its size to be changed, and for that we took the results for the  $NN$
radius obtained in the calculations of \cite{Dote:2008hw}. We found that the consideration of this reduced $NN$ size reverted into  
a larger binding of the three body system. For the  $S=0$, $I_{NN}=1$ state with ``normal'' $NN$ size we found a peak around $30$ MeV
binding and a width of about $50$ MeV, but when the reduced size considered
we found a shift of about 10 MeV to larger binding energies and practically no 
change in the width ( $ B=40$ MeV,  $ \Gamma = 50$ MeV). 
We also found a peak with smaller strength for the $S=1$, $I_{NN}=0$ (with $L_{\bar{K}N}=0$) 
configuration around 27 MeV, when the reduced size was considered. We also saw that this configuration should weakly mix
 (with $L_{\bar{K}N}=1$) with the dominant configuration $S=0$, $I_{NN}=1$. 
The results that we obtain are in line with those obtained when a chiral amplitude is used for the  $\bar{K}N$ interaction, 
or when the energy dependence of the chiral approach is used \cite{Dote:2008in,Dote:2008hw,Ikeda:2008ub}. Our results, which 
rely only upon physical on shell amplitudes,  reinforce the findings of these latter works and they shed light on why other
works could give different results, because they either force the $\Lambda(1405)$ with mass $1405$ MeV to be a bound state
of  $\bar{K}N$, or because they use an energy independent kernel to describe the $\bar{K}N$ interaction, and in any case 
because they introduce off shell effects which cannot be controlled in those approaches.  

  The simplicity of the present approach also allows for a transparent interpretation of the results,
not easy to see when one uses either a variational method or the full Faddeev equations. 
The dominance of the $S=0$, $I_{NN}=1$ channel could be anticipated once the $\bar{K}N$
amplitudes for a $N$ belonging to a cluster with $S=0$ or $S=1$ is known. This conclusion is in agreement 
with results of other methods, which were found with more laborious ways. The results obtained here with our on
shell approach with the FCA to Faddeev equations, should encourage more elaborate calculations with the 
on shell approach to Faddeev equations developed in \cite{MartinezTorres:2007sr,Khemchandani:2008rk}. Comparison with
the present
results would tell us the degree of accuracy of the present method, and with this knowledge one could venture into similar 
problems with this technically much simpler approach.

  \section{Acknowledgments}
  We thank T. Hyodo and A. Gal for useful comments and one of the authors, M. Bayar, thanks to I. Ruiz Simo for kind help. This work is partly supported by  projects FIS2006-03438 from the Ministerio de Ciencia e Innovaci\'on (Spain), FEDER funds and by the Generalitat Valenciana in the program Prometeo/2009/090.
This research is part of the European
 Community-Research Infrastructure Integrating Activity ``Study of
 Strongly Interacting Matter'' (acronym HadronPhysics2, Grant
 Agreement n. 227431) 
 under the Seventh Framework Programme of EU. M. Bayar acknowledges support through the Scientific and
 Technical Research Council (TUBITAK) BIDEP-2219 grant.


\end{document}